\documentclass{elsart}
\usepackage{amssymb}
\usepackage{graphicx}

\newcommand{\Tr}{{\mathrm Tr}}
\newcommand{\Z}{{Z \!\!\! Z}}

\newcommand{\dd}{{\mathrm d}}

\newcommand{\vx}{{\mathbf{x}}}
\newcommand{\eq}[1]{(\ref{#1})}
\newcommand{\beqn}{\begin{eqnarray}}
\newcommand{\eeqn}{\end{eqnarray}}
\def\bbbone{{\mathchoice {\rm 1\mskip-4mu l} {\rm 1\mskip-4mu l}
{\rm 1\mskip-4.5mu l} {\rm 1\mskip-5mu l}}}

\begin{document}
\runauthor{Belavin, Chernodub and Kozlov}
\begin{frontmatter}

\title{Hedgehogs in Wilson loops and phase transition in SU(2) Yang-Mills theory}

\author[ITEP]{V.\,A.~Belavin}
\author[ITEP,UU]{M.\,N.~Chernodub}
\ead{Maxim.Chernodub@itep.ru}
\author[ITEP,MSU]{I.\,E.~Kozlov}
\address[ITEP]{Institute for Theoretical and Experimental Physics,\\
B.Cheremushkinskaya 25, RU-117259, Moscow, Russia}
\address[UU]{Department of Theoretical Physics, Uppsala University,\\
P.O. Box 803, S-75108, Uppsala, Sweden}
\address[MSU]{Faculty of Physics, Moscow State University,
RU-119992, Moscow, Russia}

\begin{abstract}
We suggest that the gauge-invariant hedgehoglike structures in the
Wilson loops are physically interesting degrees of freedom in the
Yang--Mills theory. The trajectories of these ``hedgehog loops''
are closed curves corresponding to center-valued (untraced) Wilson
loops and are characterized by the center charge and winding
number. We show numerically in the SU(2) Yang--Mills theory that
the density of hedgehog structures in the thermal Wilson--Polyakov
line is very sensitive to the finite-temperature phase transition.
The (additively normalized) hedgehog line density behaves like an
order parameter: the density is almost independent of the
temperature in the confinement phase and changes substantially as
the system enters the deconfinement phase. In particular, our
results suggest that the (static) hedgehog lines may be relevant
degrees of freedom around the deconfinement transition and thus
affect evolution of the quark-gluon plasma in high-energy
heavy-ion collisions.
\end{abstract}

\date{}

\end{frontmatter}

\section{Introduction}

Topological structures in fields are as important as the fields
themselves. The well-known realization of this statement is the
three-dimensional Georgi--Glashow (GG) model, which has
topological defects called 't~Hooft--Polyakov (HP)
monopoles~\cite{ref:HP}. The presence of these defects
determine special features of the vacuum
structure~\cite{ref:compact:Polyakov}: external electrically
charged particles show linear confinement at large separations
while the magnetic fields are screened at large distances. Neither
the confinement nor the mass-gap generation are realized if the
monopoles are absent in the vacuum (like, for example,
in a case of the noncompact Abelian gauge theory).

Another textbook example is conventional superconductivity: in
the Ginzburg--Landau approach the density of the superconducting
electrons is described by the order-parameter field
corresponding to an effective field of paired electrons (Cooper pairs).
The topological structure in the Cooper-pair
field is the Abrikosov vortex, which drastically influences the
conduction and thermodynamic properties of conventional
superconductors (for example, subjected to an external magnetic
field).

The concept of topological defects in fields was
invoked to understand color confinement. This phenomenon,
which is one of the most interesting issues in QCD, is governed
by gluons dynamics described by the pure Yang--Mills (YM)
theory. The YM theories have zero-dimensional topological objects
(instantons), but these objects cannot explain the color
confinement~\cite{ref:Mitya} similarly to the GG
model~\cite{ref:compact:Polyakov}. On the other hand, the YM
theories lack topologically stable monopole-like and
vortexlike classical configurations. Nevertheless, the
confinement was suggested to be driven by condensation or,
equivalently, by proliferation of the
vortexlike~\cite{ref:vortices} and
monopole-like~\cite{ref:monopoles} nonclassical structures.
Unfortunately, these structures alone are not supported
topologically by the $SU(N)$ color symmetry of the YM theory in
the continuum space--time. Therefore, the current understanding
of color confinement is still unsatisfactory.

On the other hand, YM theory can be entirely
formulated~\cite{ref:loops} in the language of colorless composite
fields (Wilson loops), which are traces of the loop functionals
of the gluon fields. Mathematically, these loops are as good as the
gauge fields from the standpoint of describing properties (in
particular, nonperturbative) of YM theories~\cite{ref:loops}.
Having the importance of the order-parameter (Cooper-pair) field
in the Ginzburg--Landau model of superconductivity in mind,
we compare the properties of this field with the
well-known properties of the Wilson loop variable in YM theory:
\begin{itemize}
\item[{\em i.}] Expectation values:
The expectation value of the Cooper-pair field corresponds to the
density of superconducting electrons, while the vacuum
expectation value of the Wilson loop in YM theory is related
to the interquark interaction. Both expectation values are
order parameters (nonlocal in the case of the Wilson loop): the
superconducting phase in ordinary superconductors is
determined by a nonzero value of the Cooper-pair field, while the
confining phase in the YM theories is manifested by an area-like
asymptotic behavior of the Wilson loop.
\item[{\em ii.}] Correlation functions:
The correlation function of the Cooper-pair fields determines the
type and spectrum of the superconducting medium, while
correlators of the Wilson loops provide information about the
glueball spectrum of the strongly interacting medium, the
intermeson interactions, {\it etc}.
\item[{\em iii.}] Topology: the topological structure in
the Cooper-pair field is the celebrated Abrikosov vortex.
What is an analog of this structure in the gluon fields of YM theory
in the language of the Wilson loop variable? The answer to this
question may be important for understanding color confinement
because the Wilson loop operator itself is intimately related to
this important phenomenon. Moreover, the mentioned structure in the Wilson
loop must be a gauge-invariant quantity like the confinement phenomenon.
\end{itemize}

General arguments were given in~\cite{ref:objects} to identify the
defect-like trajectories in the Yang-Mills theory with the closed paths
in space for which the (untraced) Wilson loop takes its value in the
center of the color group. The main argument in~\cite{ref:objects}
is that the gluon fields in the vicinity of such ``extremal'' loops are
similar to the gauge-invariant hedgehoglike structures, which resemble the
HP monopoles in the GG model. Below we refer to these structures
as to the ``hedgehog loops''. Developing the idea further, we
here investigate properties of the hedgehogs loops at a finite
temperature. In particular, the static (thermal) hedgehoglike
structures -- which are also called below as the ``hedgehog lines'' --
should definitely arise in an effective theory of Polyakov
lines~\cite{ref:Pisarski1}, which is proposed to describe basic
features of the finite-temperature QCD transition. Moreover, the
thermalization properties of the Polyakov line
``condensate''~\cite{ref:Pisarski1} give a qualitatively
encouraging insight regarding particle production in high-energy
heavy-ion collisions~\cite{ref:Pisarski2}. As we show below, the
density of the hedgehog lines in the deconfinement
phase increases rapidly as temperature increases. This observation
suggests that the hedgehog lines, corresponding to the extremal values
in the Polyakov line variables, may be important degrees of freedom in
quark-gluon states created in high-energy heavy-ion collisions.

The content of this paper is as follows. In
Sec.~\ref{sec:hedgehogs}, we define the hedgehog structure
in the Wilson loop. Section~\ref{sec:confinement} is
devoted to a general discussion of hedgehog loop properties.
In Sec.~\ref{sec:density}, we derive an expression
for the density of the static hedgehogs (the hedgehog lines) as
a function of the trace of the Polyakov line. We present the results
of our numerical simulations in Sec.~\ref{sec:lattice} and
state conclusions in the last section.

\section{Center-valued Wilson loops as hedgehog loops}
\label{sec:hedgehogs}

The hedgehogs in the Wilson loops are similar to the HP monopoles in the GG model~\cite{ref:objects}.
In this model the triplet Higgs field $\hat \Phi\equiv \frac{1}{2} \Phi^a \sigma^a$
of the static HP monopoles ~\cite{ref:HP} vanishes at the center of the HP monopole $\vx=\vx_0$,
\beqn
\hat\Phi(\vx_0) & = & 0\,.
\label{eq:phi}
\eeqn
Moreover, in spatial vicinity of a HP monopole the Higgs field has a generic hedgehog
structure in the in a gauge where the gauge field~$A_\mu$ is regular :
\beqn
\hat \Phi(\vx) & = & \frac{\sigma^a}{2} Y^{ai}(x_0) \cdot (x-x_0)^i + O\Bigl((\vx-\vx_0)^2\Bigr)\,.
\label{eq:phi:exp}
\eeqn
Hedgehog structure~\eq{eq:phi:exp} inevitably implies that
Higgs field~\eq{eq:phi} vanishes, while the converse is false in general case.
On the other hand, if the triplet field $\Phi$ vanishes
at the point $\vx_0$, then a general Taylor expansion of the field
in the vicinity of this point starts from linear terms, and
expansion~\eq{eq:phi:exp} therefore corresponds to a generic
situation. Note that the matrix $Y$ is assumed to be nondegenerate in a general case,
$${\mathrm{det}} \, Y \neq 0\,.$$

In YM theory,a hedgehog structure can be entirely defined in terms of
Wilson-loop variables~\cite{ref:objects}. We consider an untraced Wilson loop beginning and
ending at the point $x_0$ on the closed loop~${\mathcal C}$,
\beqn
W_{\mathcal C}(x_0) = P\,\exp\Bigl\{i g\, \oint_{{\mathcal C}} \dd x_\mu\, A_\mu(x) \Bigr\}\,.
\label{ref:W}
\eeqn
This quantity transforms locally as an adjoint operator
with respect to the SU(2) color symmetry,
$W_{\mathcal C}(\vx_0) \to \Omega^\dagger(\vx_0) W_{\mathcal C}(\vx_0) \Omega(\vx_0)$.
To improve the analogy with the triplet Higgs field $\hat \Phi$,
we subtract the singlet part from~\eq{ref:W}
\beqn
\hat \Gamma_{\mathcal C}(x_0) = W_{\mathcal C}(x_0) - \bbbone \cdot \frac{1}{2}\, {\mathrm{Tr}}\, W_{\mathcal C}(x_0)\,.
\label{ref:Gamma}
\eeqn
This is a traceless adjoint operator similar to the field $\hat \Phi$.

The next step is to associate the triplet part $\hat \Gamma_{\mathcal C}$
of Wilson loop~\eq{ref:Gamma} with the triplet Higgs field
$\hat \Phi(x)$ in the GG model. To uncover a link between
the hedgehog structures in the fields of the GG model and in the extremal loops
of the YM theory, we notice the following~\cite{ref:objects}:
\begin{enumerate}
\item[{\em i.}] The Higgs field vanishes in all points $x$ belonging to the
HP monopole trajectory, Eq.~\eq{eq:phi}. Similarly, one could expect that the
Wilson loop evaluated on the hedgehog loop ${\mathcal C}$ belongs to the center of the gauge
group, $W_{\mathcal C} = \pm \bbbone$. In turn, this implies that $\Gamma_{\mathcal C}$ vanishes on
the hedgehog loop ${\mathcal C}$:
\beqn
W_{\mathcal C} \in {\mathbb Z}_2 \qquad \Leftrightarrow \qquad \Gamma_{\mathcal C}=0\,.
\label{eq:condition:1}
\eeqn
We intentionally omit the argument $x_0$ in
the loop variables in Eq.~\eq{eq:condition:1} because
condition~\eq{eq:condition:1} turns out to be independent of the choice
of the reference point $x_0$ on the trajectory ${\mathcal C}$~\cite{ref:objects}.
In other words, if $W_{\mathcal C}(x_1)\in{\mathbb Z}_2$ for at least one point
$x_1\in{\mathcal C}$, then $W_{\mathcal C}(x_0)\in{\mathbb Z}_2$ for all points $x_0 \in {\mathcal C}$.
This relation reflects self-consistency of our definition of the hedgehog loop.

In order to illustrate the self-consistency, let us consider two Wilson loops lying on the same contour ${\mathcal C}$ but
open at two different points $x_0$ and $x_1$. Both points are arbitrary.
Then the corresponding Wilson loops~\eq{ref:W} are related to each other by the adjoint transformation:
$$
W_{\mathcal C}(x_1) = U^\dagger(x_1,x_0)\,W_{\mathcal C}(x_0)\, U(x_1,x_0)\,,
$$
with
$$
U(x_1,x_0) =
P\,\exp\Bigl\{i g\, \oint^{x_1}_{x_0} \dd x_\mu\, A_\mu(x) \Bigr\} \in SU(2)\,.
$$
Obviously, if $W_{{\mathcal C}}(x_0) \in \Z_2$ then $W_{{\mathcal C}}(x_1) \in \Z_2$ as well, and,
consequently, $\Gamma_{{\mathcal C}}(x_1) \equiv 0$. Thus the condition~\eq{eq:condition:1} is
insensitive to the reference point~$x_0$.

One can alternatively state that the independence of the definition of the hedgehog on the reference
point $x_0$ manifests a conservation of the central charge $z$, which is defined in point~{\em iv} below.

\item[{\em ii.}] The triplet fields around the HP monopole
have the structure of hedgehog~\eq{eq:phi:exp}. To demonstrate that the
center-valued Wilson loop variables have something to do with the hedgehog structure,
we infinitesimally deform the contour ${\mathcal C}\to{\mathcal C}+\delta{\mathcal C}$
in the vicinity of the point $x_0$. This shift also
relocates the reference point, $x_0 \to x = x_0 + \delta x$
(see Fig.~\ref{fig:contour:general}).
\begin{figure}[!thb]
\vskip 5mm
\begin{center}
\includegraphics[scale=1.4,clip=true]{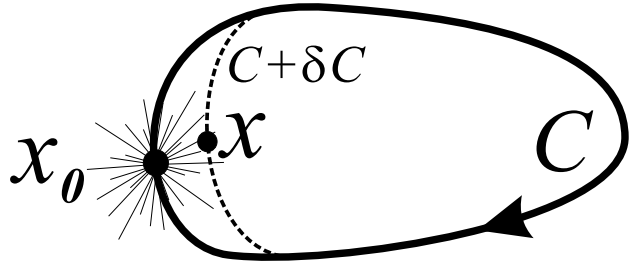}
\end{center}
\caption{The Wilson loop ${\mathcal C}$ (solid line) and its
variation (dotted line).}
\label{fig:contour:general}
\vskip 3mm
\end{figure}
Without loss of generality, we assume that a tangent vector to the
loop ${\mathcal C}$ at the point $x_0$ is aimed in the time
direction. A physically distinguishable shift $\delta x$ should be
tangential (spacelike) to the contour ${\mathcal C}$ because
longitudinal (timelike) shifts of the point $x_0$ leave the
functional $\Gamma_{\mathcal C}$ invariant.

Then the deformed loop $\Gamma_{\mathcal C}(x)$, in general case, should have
a spatial hedgehoglike structure in the vicinity of the zero point $x_0$,
\beqn
\Gamma_{\mathcal C}(x) = \frac{\sigma^a}{2} \, {\mathcal Y}^{ai}_{\mathcal C}(x_0) \cdot
{\bigl(x - x_0\bigr)}^i + O\Bigl(\bigl(x - x_0\bigr)^2\Bigr)\,.
\label{eq:condition:2}
\eeqn
This structure is similar to that of Eq.~\eq{eq:phi:exp} with
only one exception: the definition of the matrix ${\mathcal Y}_{\mathcal C}$
involves the path derivative instead of the usual derivative,
\beqn
{\mathcal Y}^{ai}_{\mathcal C}(x) = \frac{\delta \Gamma_{\mathcal C}^a(x)}{\delta x^i}{\Biggl|}_{x \to x_0}\,, \qquad a,i=1,2,3\,.
\label{eq:Y:variation}
\eeqn
which defines a change of the functional $\Gamma_{\mathcal C}$ under an infinitesimal change of the contour
${\mathcal C} \to {\mathcal C} + \delta C$.

It is worth stressing that a loop at which the Wilson loop takes
its value in the center of the gauge group correspond, in general
case, to a hedgehog loop. This statement may not be true in a
degenerate case, ${\mathrm{det}} {\mathcal Y}^{ai}_{\mathcal C}(x) = 0$,
which is realized, for example, in a trivial vacuum, $A^a_\mu=0$. The
degenerate case is nongeneric and thus is statistically
suppressed.

Note that the matrix~\eq{eq:Y:variation} can also be written in a explicitly Lorentz--invariant form,
$$
{\mathcal Y}^{a\nu}_{\mathcal C}(x) \propto m^\mu_{\mathcal C}\, F^a_{\mu \nu}\,,
$$
where $m^\mu_{\mathcal Y}(x) = \dot{x}^\mu_{\mathcal Y}/|\dot{x}_{\mathcal Y}|$
is a vector which is tangent to the contour ${\mathcal Y}$
at the point $x$. The contour ${\mathcal Y}$ is parameterized by the vector
function ${\bar x}^\mu_{\mathcal Y}$ of the variable $\tau$ and
$\dot{x}^\mu_{\mathcal Y} \equiv \partial {\bar x}^\mu_{\mathcal Y}(\tau)/\partial \tau$.

\item[{\em iii.}] The location $x_0$ of the HP monopole
is a gauge-invariant quantity.
Similarly, the trajectory ${\mathcal C}$ of the hedgehog loop in the
Wilson loop defined by condition~\eq{eq:condition:1} is gauge
invariant because the Wilson loop eigenvalues are gauge invariant.

\item[{\em iv.}] The $SU(2)$ HP monopoles are classified
by the integer-valued winding number
$n \in \pi_2(S^2) \simeq {\mathbb Z}$ corresponding the number of
times the vacuum manifold $S^2$ is covered by a single turn around
the two-dimensional boundary $\partial D_3 \simeq S^2$ of the
three-dimensional volume $D_3$ containing the HP monopole.
Hedgehog structure~\eq{eq:phi:exp} with $Y^{ai} \propto
\delta^{ai}$ corresponds to the $n{=}1$ HP monopole in the GG model.
In YM theory, the hedgehog loop is classified by the
{\it pair} of numbers $(z,n)$, where $z = \pm 1$ is the central
element of the Wilson loop $W_{\mathcal C} \in Z_2$ determined by
Eq.~\eq{eq:condition:1} and $n$ is the winding number associated
with hedgehogs structure~\eq{eq:condition:2}. Below, we call
$z$ the ``center charge'' of the hedgehog loop.
A hedgehog line, which is a particular case of a general YM hedgehog in
a Wilson loop, is also classified by the two charges.

\item[{\em v.}] The gluonic fields around a static hedgehog (or, equivalently,
a thermal hedgehog line) share similarity with a field of a monopole in the
axial gauge studied in~\cite{ref19}. However, in general,
the hedgehog singularity has something to do with electric fields
rather than with the magnetic ones. Indeed in the case of a static
trajectory the matrix ${\mathcal Y}^{a\nu}$ becomes equal to the
chromoelectric field, $E^a_i \equiv F^a_{4i}$, as one can deduce from
Eq.~\eq{eq:Y:variation} by an explicit calculation.

\item[{\em vi.}] Because the hedgehog loops are closed, the
hedgehog charges must be conserved.

\end{enumerate}

\section{Hedgehog loops and color confinement}
\label{sec:confinement}

As noted in the introduction, a close relation between the
hedgehog loops and the color confinement problem can
be understood intuitively on general grounds because the vacuum
expectation value of the Wilson loops recognizes the color
confinement. The hedgehog loops, corresponding to the extremal-valued
Wilson loops, should also be sensitive to the
color confinement. In this section, we argue in support of this
suggestion.

In conventional superconductivity, Abrikosov vortices are
singularities in the superconducting condensate (i.e., in
the Cooper-pair field). In the core of the Abrikosov vortices, the
superconductivity is broken, and the normal metal state is
restored. It is well known that the weaker the Cooper-pair
condensate is, the lighter (less tense) the Abrikosov vortices are. As
temperature increases, the condensate weakens, and nucleation
of the vortices due to thermal fluctuations strengthens. Thus,
the higher the temperature is, the denser the (thermal) vortex
ensemble should be.

It can be expected in the YM theory that the density of hedgehog
loops is also sensitive to the phase transition.
In the Euclidean formulation of the field theory, the
temperature direction of the system volume is compactified to a
circle whose circumference is equal to the inverse temperature.
The order parameter of the phase transition is the vacuum
expectation value of the (trace of the) Polyakov line,
\beqn
\hat L(\vx) = P\,\exp\Bigl\{i g\, \oint\limits_0^{1/T} \dd x_4\, A_4(\vx,x_4) \Bigr\}\,,
\label{ref:L}
\eeqn
which is static Wilson loop~\eq{ref:W} closed through the boundary of
the compactified (temperature) direction (we here define
$L \equiv \frac{1}{2} {\mathrm{Tr}} \hat L$). Functional~\eq{ref:L},
also called the thermal Wilson line, is a basic variable
in an effective theory proposed in~\cite{ref:Pisarski1} to describe
the properties of the finite-temperature QCD transition.
In the confinement phase, the expectation value of the Polyakov line is zero, $\langle L \rangle \equiv e^{- T F_q}= 0$,
indicating that the free energy of a single quark becomes infinite, $F_q
\to \infty$. In the deconfinement phase, the Polyakov line has a
nonzero expectation value, $\langle L \rangle \neq 0$, and the
quarks are no longer confined.

It can be expected that static hedgehog lines ({\it i.e.}, objects
corresponding to the hedgehog structures in the Polyakov lines) are
sensitive to the phase transition because the Polyakov line is the
order parameter of this transition. In fact, the following analogy
between the Cooper-pair condensate and the Polyakov line can be
established. Both quantities are the order parameters, and
the singularities in the Cooper pair condensate and the extremal values of
the Polyakov lines should therefore be physically relevant degrees of freedom in
the corresponding theories. As we show in the following sections, this
suggestion is indeed correct. We note that this analogy is only partial
because the core of the Abrikosov vortex corresponds to the
unbroken vacuum while the vacuum in the core of the hedgehog loop
in the Yang-Mills theory is maximally broken.

The behavior of the static hedgehog lines can
be found qualitatively using symmetry arguments. The
confinement--deconfinement phase transition is accompanied by
breaking the global center symmetry in the deconfinement phase.
This center symmetry is realized in the case of the $SU(2)$
Yang--Mills theory by the transformations $L(x) \to z\, L(x)$,
where the factor $z = \pm1$ belongs to the center of the gauge
group, $z \in \Z_2$. In the confinement phase, the center symmetry
is unbroken, and the effective potential on the Polyakov variable,
$V_{\mathrm{eff}}(L) = b_2 L^2/2 + b_4 L^4/4 + \dots$ with $b_{2,4}>0$,
has a minimum at the symmetric vacuum $\langle L \rangle=0$,
which is far away from the central values $L_\pm =\pm1$.
Therefore, the density of the hedgehogs lines should be small in
the confinement phase.

In the deconfinement phase, the center symmetry is broken by the
vacuum, and (as the temperature passes the critical value $T = T_c$)
the single minimum of the effective potential splits into two ground states,
$L_{\min} = \pm \sqrt{-b_2/b_4}$ with $b_2<0$,
one of which is picked up by the vacuum $\langle L \rangle$. As the
temperature increases further, the vacuum state approaches one of
the central values $L_{\min} \to L_\pm = \pm 1$. Therefore, we
expect to observe an increase in the density of static
hedgehog lines characterized by the central charge $z$ closest
to the vacuum state $\langle L \rangle$. The density of
static hedgehog lines with the larger value of $|\langle L \rangle - z|$
should decrease in the deconfinement phase. The density of
such intolerable hedgehog lines for $T>T_c$ should be lower than
the density in the confinement phase. It is clear that at the
critical temperature, the hedgehog line density should split into two
branches.

The qualitative behavior of the spatial Wilson loops is known to
be rather insensitive to the deconfinement transition: the area
law for the spatial Wilson loops is observed in both the
confinement and deconfinement phases. Therefore, we do not
expect a drastic change in the density of spatial hedgehog loops as
the YM system goes through the phase transition.

\section{Density of static hedgehog lines}
\label{sec:density}

To go beyond the qualitative analysis presented in the
preceding section, we would like to simulate the
YM theory on the lattice numerically. But we immediately notice that
the definition of the hedgehog loop~(see~\ref{eq:condition:1}
and~\ref{eq:condition:2}) is rather inconvenient from the standpoint
of numerical simulations.
Indeed, the trajectory of the hedgehog structure in the Wilson
loops cannot be determined locally as can be done in the case
of the Abrikosov vortices in the Ginzburg--Landau model of
superconductivity or in the case of the HP monopoles in the GG model.
To check whether a hedgehog loop passes through a particular point $x$
in the space--time for one particular gauge-field configuration,
we should analyze the Wilson
loops $W_{\mathcal C}$ corresponding to {\it all} closed loops ${\mathcal C}$ that
pass through the point $x\ni{\mathcal C}$. Obviously, it is
difficult to realize this definition in numerical simulations
of the YM theory on the lattice.

But at a finite temperature, static hedgehog lines are the most interesting
because their density should show a noticeable change
(in contrast to the spatial hedgehog loops) at the phase transition.
In a continuum limit, it is easy to check whether a static
hedgehog line passes through the particular point $x$ of the space.
For this, we should evaluate (untraced) Polyakov line~\eq{ref:L}
and then check whether it belongs to the center of the group.
In the $SU(2)$ YM model, the normalized trace of the Polyakov line
on a static hedgehog line should be equal to $\pm 1$.
In this section, we derive an analytic expression for
the (absolute value of the) density of a static hedgehog line
as a function of the {\it traced} Polyakov line $\rho =\rho[L]$.

Before proceeding further, we point out that in the static limit,
hedgehog lines are similar to Abelian monopoles
in the Polyakov and the Axial Abelian gauges
of the SU(2) YM theory~\cite{ref:Suzuki:Polyakov,ref19,ref:Polyakov:other}.
Because the Abelian monopoles in the Abelian Polyakov gauge are always
static in the continuum limit, they are not related to the
confinement of static quarks, but they still may be responsible
for the spatial string tension at high temperatures~\cite{Chernodub:2003mm}.
Moreover, such monopoles are related to the topological charge
in the YM theory~\cite{ref19,ref:Polyakov:other}.
The Polyakov-loop variable can also be used to find
(static) monopole constituents in
physically interesting topologically nontrivial
configurations~\cite{ref:Falk}. We note that the density of
static hedgehog lines (as well as the Polyakov line expectation value)
is dual in a sense to the Abelian monopole condensate because the
density of the hedgehog lines is expected to be an order parameter of the
finite-temperature phase transition while the monopole condensate
is a disorder parameter~\cite{ref:independence}.

\subsection{Density of static hedgehog lines in continuum}

In general, the untraced Polyakov line can be expanded in the quaternion basis:
\beqn
\hat L(\vx,x_4) = L(\vx)\cdot \bbbone + i (\vec \Gamma(\vx,x_4) \cdot \vec \sigma)\,,
\label{representationL}
\eeqn
where the singlet part of $L$ is independent of $x_4$ and the components
of the vector $\vec\Gamma$ describe the gauge-variant triplet part.
If a hedgehog line passes through the point $\vx_0$, then $L(\vx) = \pm 1$ and
$\vec \Gamma(\vx_0) = 0$. Therefore, the triplet part in a neighborhood
of the point $\vec x_0$ can be expanded as
\beqn
\Gamma^a(\vx) = Y^{ai}(\vx_0) \cdot (x - x_0)^i + O\Bigl((\vx - \vx_0)^2\Bigr)\,,
\label{eq:taylor}
\eeqn
where ${\mathrm {det}}\, Y^{ab} (\vx_0) \neq 0$ in general case.
We do not consider the rare degenerate case ${\mathrm {det}}\, Y^{ab} (\vx_0) = 0$
here. Therefore the density of the static hedgehog line corresponds to the density
of the points, where the Polyakov line takes its value in the center of the group.
We note that Eq.~\eq{eq:taylor} is similar to Eqs.~\eq{eq:phi:exp} and~\eq{eq:condition:2}.

The density of the hedgehog lines is expressed as
\beqn
\rho = \gamma \, \delta (\vec \Gamma^2) \equiv \frac{\gamma}{2 |\vec \Gamma|} \cdot \delta(|\vec \Gamma|)\,,
\label{eq:gamma1}
\eeqn
where $\gamma$ is an unknown factor to be determined below.
Because the quantity $\vec \Gamma$ is dimensionless,
the dimension of the factor $\gamma$ should be [$mass^3$].

Using the identity $\delta(r) = {4 \pi r^2} \delta^{(3)}(\vec r)$
and Eq.~\eq{eq:gamma1}, we obtain $\rho[L]= 2 \pi \gamma |\vec \Gamma|
\, \delta^{(3)}(\vec \Gamma)$. Integrating this identity
in the infinitesimally small volume $v_{\vx_0}$ with $\vx_0 \ni v_{\vx_0}$
gives the self-consistency condition, which allows us to determine the factor $\gamma$
in Eq.~\eq{eq:gamma1}:
\beqn
1 = \int\limits_{v_{\vx_0}} {\mathrm{d}} \vx \,\rho(\vx) = 2 \pi \int\limits_{v_{\vx_0}}
{\mathrm{d}} \vx \, \gamma(\vx) |\vec \Phi(\vx)| \delta^{(3)}(\vec \Phi(\vx))\,.
\label{conjunction}
\eeqn
Changing the integration variables from ${\mathrm d} \vx$
to ${\mathrm d} \vec \Gamma$, we obtain the Jacobian factor,
\beqn
{\mathrm d} \vx = \frac{1}{|{\det \frac{\partial \Gamma^a}{\partial x^b} }|} \, {\mathrm{d}} \vec \Gamma
= \frac{1}{|\det Y^{ab}(\vx_0)|} \, {\mathrm{d}} \vec \Gamma\,.
\label{mera}
\eeqn
Substituting Eq.~\eq{mera} in Eq.~\eq{conjunction}, we obtain
\beqn
\gamma = \frac{\det |Y^{ab}|}{2 \pi |\vec \Gamma|} \equiv
\frac{1}{4\pi \sqrt{2(1-L^2)}} \,
\sqrt{\det \frac{\partial^2 L^2}{\partial x^i \partial x^j}}\,,
\label{gammasearch1}
\eeqn
where we use the identity $|\vec \Gamma| = \sqrt{1 - L^2}$ and the relation
\beqn
\frac{\partial^2 L^2(\vx)}{\partial x^i \partial x^j} {\Bigl|}_{\vx=\vx_0} =
2\, Y^{ai}(\vx_0) \, Y^{aj}(\vx_0)\,,
\eeqn
which follows from Eq.~\eq{eq:taylor}. Noticing that
${\mathrm{det}} [2\, Y^T(\vx_0) \, Y(\vx_0)] \equiv 8 ({\mathrm {det}} Y)^2$,
we obtain Eq.~\eq{gammasearch1}.
Substituting Eq.~\eq{gammasearch1} in Eq.~\eq{eq:gamma1},
we finally write the expression for the monopole density:
\beqn
\rho[L(\vx)] = \frac{1}{16} \frac{1}{\sqrt{2(1-L^2(\vx))}}
\, \sqrt{\det {\Bigl|\Bigl|\frac{\partial^2}{\partial x^i \partial x^j} L^2(\vx)\Bigr|\Bigr|}_{i,j}}
\, \delta\left(1-L^2(\vx)\right)\,,
\label{eq:object:density}
\eeqn
where $L(\vx)$ is the Polyakov line evaluated at the point $\vx$.
The determinant is taken over the three spatial indices $i,j = 1,2,3$.

There are three important properties of formula~\eq{eq:object:density}.
First, this expression depends only on the singlet part (trace)
of the Polyakov line and is independent on the triplet part.

Second, Eq.~\eq{eq:object:density} corresponds to the absolute
density of hedgehog lines with a particular value of the central
charge $z$. As mentioned, the hedgehog lines are described by
the winding number $n$ and the center charge $z$.
Equation~\eq{eq:object:density} treats hedgehog lines with $\pm n$
winding numbers on equal footing while discriminating between
different center charges $z$. Knowledge of the absolute density is
enough for our exploratory study reported below. The information
about the sign of the hedgehog charge is given by the sign of the
determinant of the matrix $Y$, which can be computed explicitly if
the triplet part of the (untraced) Polyakov line is calculated.

Third, Eq.~\eq{eq:object:density} treats the loop
variations~\eq{eq:Y:variation} as ordinary variations because the
object of our interest is static by definition.
Equation~\eq{eq:object:density} can be generalized to the
nonstatic case by properly treating the loop variation
discussed in Sec.~\ref{sec:hedgehogs}.

\subsection{Density of static hedgehog lines on the lattice}

\begin{figure}[htb]
\vskip 3mm
\centerline{\includegraphics[width=130mm,angle=0]{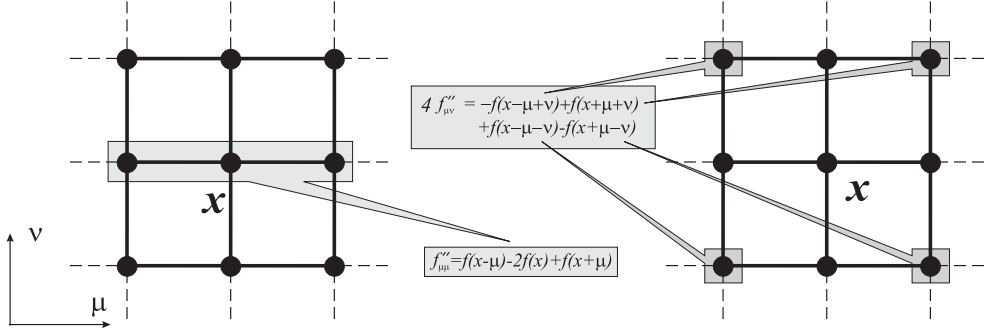}}
\caption{The lattice double derivatives
$f''_{\mu\nu}(\vx) \equiv \Delta_\mu \Delta_\nu \, f(\vx)$
at the point $\vx$.}
\label{fig:derivatives}
\end{figure}

The density of hedgehog lines in the continuum
space--time is given by Eq.~\eq{eq:object:density}.
To obtain the corresponding equation on the lattice,
we discretize the double derivatives in Eq.~\eq{eq:object:density}
straightforwardly:
\beqn
\frac{\partial^2}{\partial x^\mu \partial x^\mu} f(\vx) \to \Delta_\mu \Delta_\mu \, f(\vx)
& = & f(\vx-\hat\mu)- 2 f(\vx) + f(\vx+\hat\mu)\,, \label{eq:delta:mu}\\
\frac{\partial^2}{\partial x^\mu \partial x^\nu} f(\vx) \to \Delta_\mu \Delta_\nu \, f(\vx)
& = & \frac{1}{4} \sum_{P_\mu,P_\nu=\pm 1}
P_\mu P_\nu \, f(\vx+P_\mu\cdot \hat\mu+P_\nu\cdot \hat\nu) \,,
\eeqn
where no sum over $\mu$ in Eq.~\eq{eq:delta:mu} is implied.
An illustration of the double derivatives is provided in Fig.~\ref{fig:derivatives}.

The lattice version of density~\eq{eq:object:density}
can be expressed in the form
\beqn
\rho\left[L(\vx)\right] = \rho_+\left[L(\vx)\right] + \rho_-\left[L(\vx)\right]\,,
\label{eq:object:density:evaluate:1}
\eeqn
where the contributions of the positive and negative center loops
to the total density can {\it formally} be written as
\beqn
\rho_\pm\left[L(\vx)\right] = d_\pm\left[L(\vx)\right] \delta\Bigl(1 \mp L(\vx)\Bigr)\,.
\label{eq:object:density:evaluate:2}
\eeqn
The weights of the separate contributions
of the positive and negative central loops are
\beqn
d_\pm\left[L(\vx)\right] =
\frac{1}{64 \, \sqrt{1 \mp L(\vx)}} \sqrt{\det {\Bigl|\Bigl|\Delta_i \Delta_j L^2(\vx)\Bigr|\Bigr|}_{i,j}}\,,
\label{eq:object:density:evaluate:3}
\eeqn

Equation~\eq{eq:object:density:evaluate:2} is not suitable for use in the lattice simulations.
From the practical standpoint, instead of Eq.~\eq{eq:object:density:evaluate:2} it is convenient to evaluate
the distributions
\beqn
\varrho_\pm(L_0) = \left\langle d_\pm\left[L(\vx)\right] \delta\Bigl(L_0 - L(\vx)\Bigr) \right\rangle\,.
\label{eq:rho:distribution}
\eeqn
Then the expectation values
of monopole densities~\eq{eq:object:density:evaluate:1}
are given by the limits
\beqn
\langle\rho_{\pm}\rangle = \lim_{L_0\to \pm 1} \varrho_\pm(L_0)\,.
\label{eq:rho:average}
\eeqn

\section{Static hedgehog lines from lattice simulations}
\label{sec:lattice}

To investigate the static hedgehog lines numerically, we
simulated the SU(2) gauge theory with the standard Wilson action on
the asymmetric ($L_s^3 \times L_t = 20^3 \times 4$) lattice with
periodic boundary conditions in all directions. We used 100
independent configurations for each value of the gauge coupling,
which ranges from $\beta=2.2$ to $\beta=2.5$. The SU(2) gauge
theory has the second-order phase transition at
$\beta\approx 2.30$ for this lattice geometry~\cite{Engels:1989fz}.

Before presenting the results for the center-valued loops,
we first discuss some well-known properties of the Polyakov lines.
The lattice Polyakov line is
\beqn
P_{\vx}[U] = \frac{1}{2}\, \Tr \prod_{x_4=0}^{L_t-1} \,
U_4(\vx,x_4)\,,
\label{eq:Polyakov:lop:definition}
\eeqn
where $U_{x,\mu}$ is the lattice gauge field. The distribution of the
Polyakov lines,
\beqn
D(L) = \langle \delta(P_{\vx}[U] - L)
\rangle\,,
\label{eq:Polyakov:distribution:definition}
\eeqn
is symmetric in the confinement phase and asymmetric in the
deconfinement phase in a sufficiently large system volume with
periodic boundary conditions~\cite{ref:Kennedy}. These properties
reflect unbroken and broken $\Z_2$ central symmetry, $L \to - L$,
in the respective confinement and deconfinement phases.

\begin{figure}[htb]
\begin{tabular}{cc}
\includegraphics[width=65mm,angle=0,clip=true]{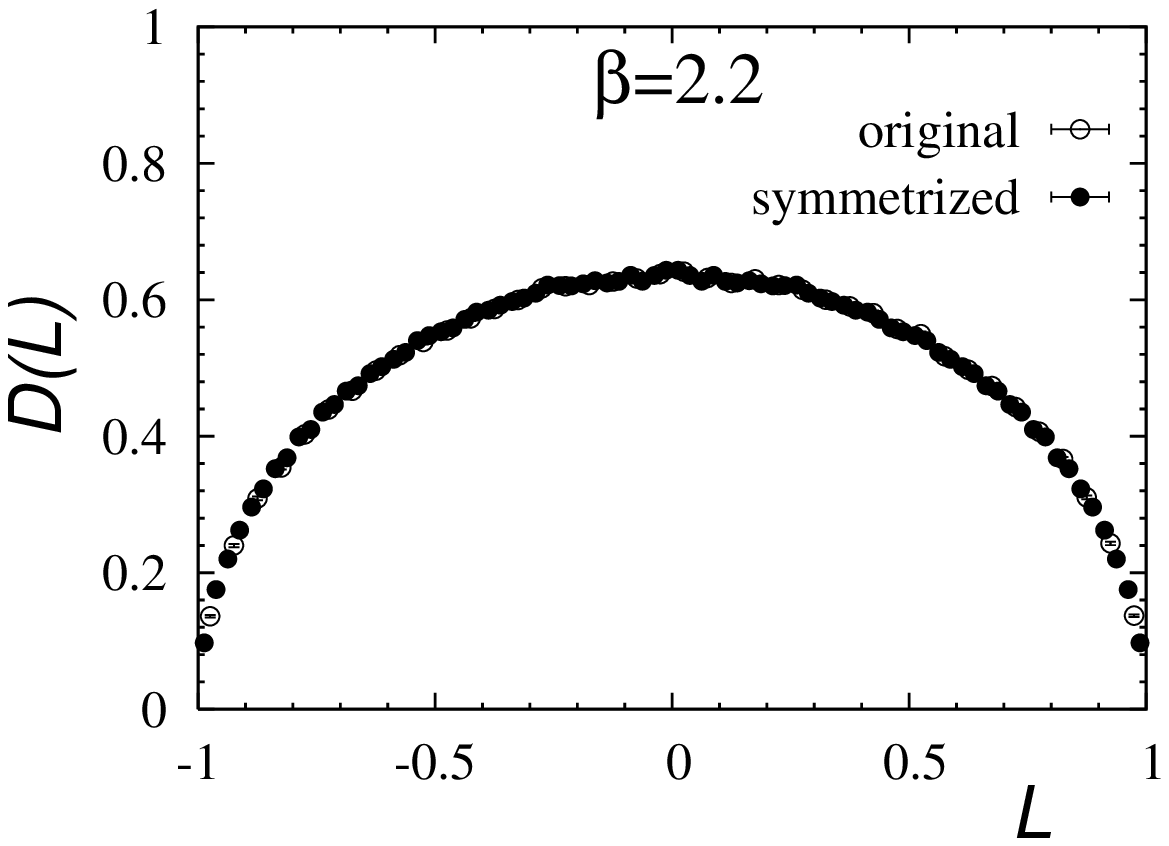} &
\includegraphics[width=65mm,angle=0,clip=true]{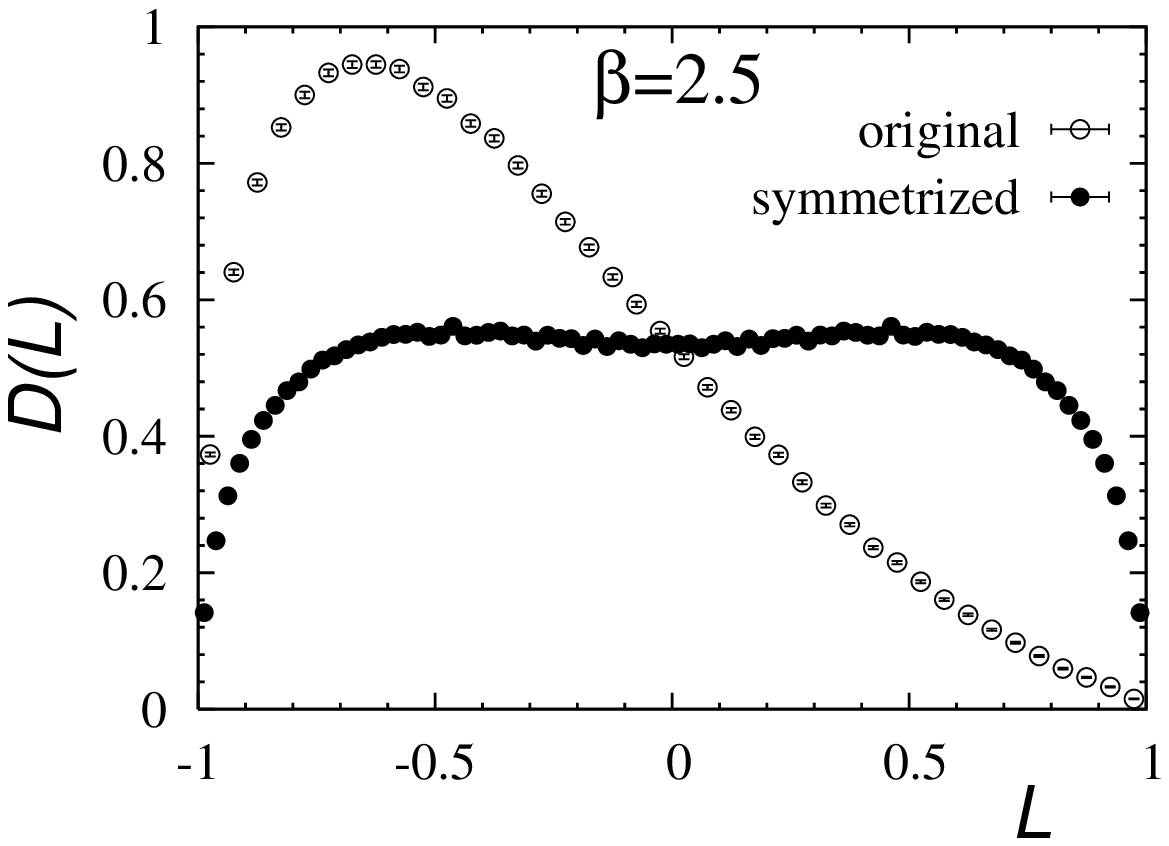} \\
(a) & (b)
\end{tabular}
\caption{Typical distributions~\eq{eq:Polyakov:distribution:definition}
of Polyakov lines~\eq{eq:Polyakov:lop:definition} in the confinement
(a) and deconfinement (b) phases are shown by open circles. The
filled circles correspond to $L \leftrightarrow - L$ symmetrized
distributions.}
\label{fig:typical:distributions}
\end{figure}

Examples of typical distributions obtained for our gluon
configurations\footnote{The Polyakov line distributions
can be modified by specific boundary conditions (for example, twisted)
in a finite volume. The properties of Polyakov lines on the torus
with twisted boundary conditions are discussed in Ref.~\cite{ref:MMP}.}
are shown in Fig.~\ref{fig:typical:distributions}.
We also show the distribution that is symmetrized under the discrete
transformation $L \to - L$, which may be associated with jumps of the system
from one $\Z_2$ minimum to the other.
In the deconfinement phase, the symmetric distribution
has two maximums, which are known to become more pronounced as the volume of
the system increases. These maximums in distributions correspond to
minimums of the effective potential for the Polyakov variables discussed
in Sec.~\ref{sec:confinement}.

\begin{figure}[htb]
\begin{tabular}{cc}
\includegraphics[width=65mm,angle=0,clip=true]{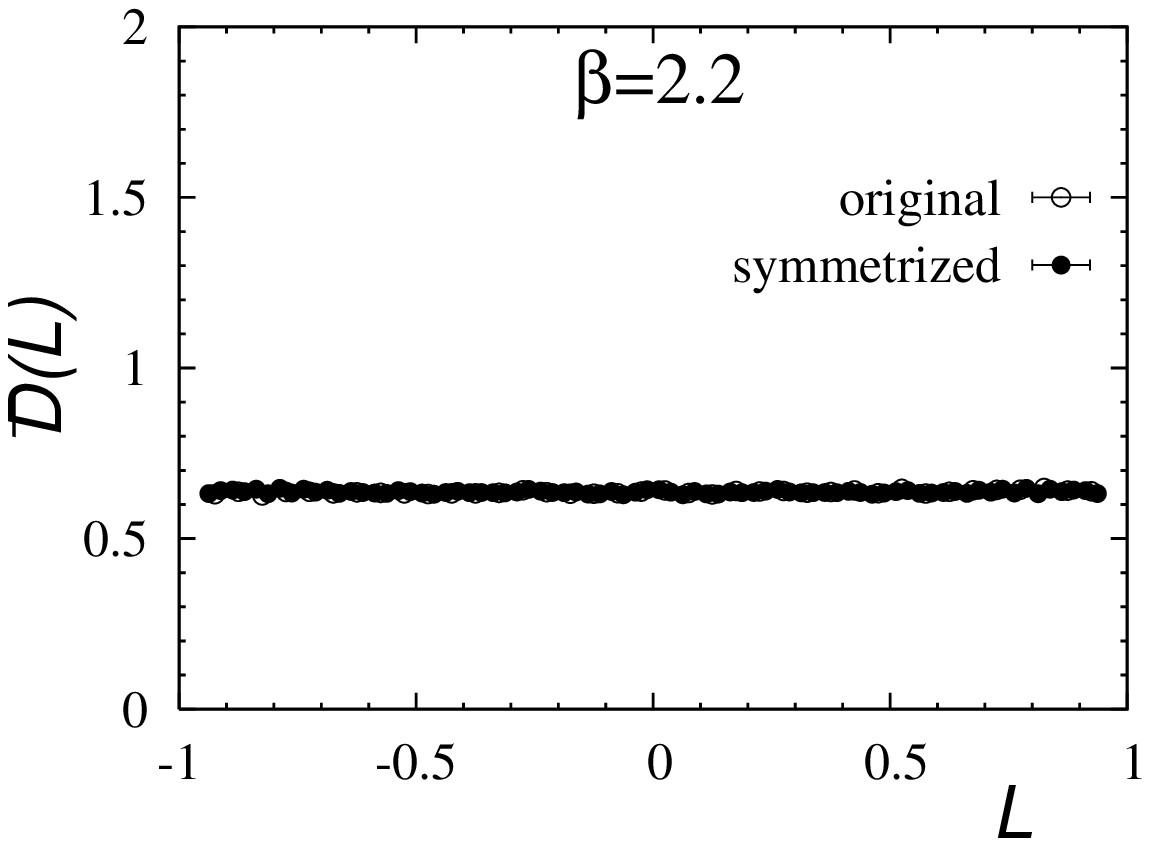} &
\includegraphics[width=65mm,angle=0,clip=true]{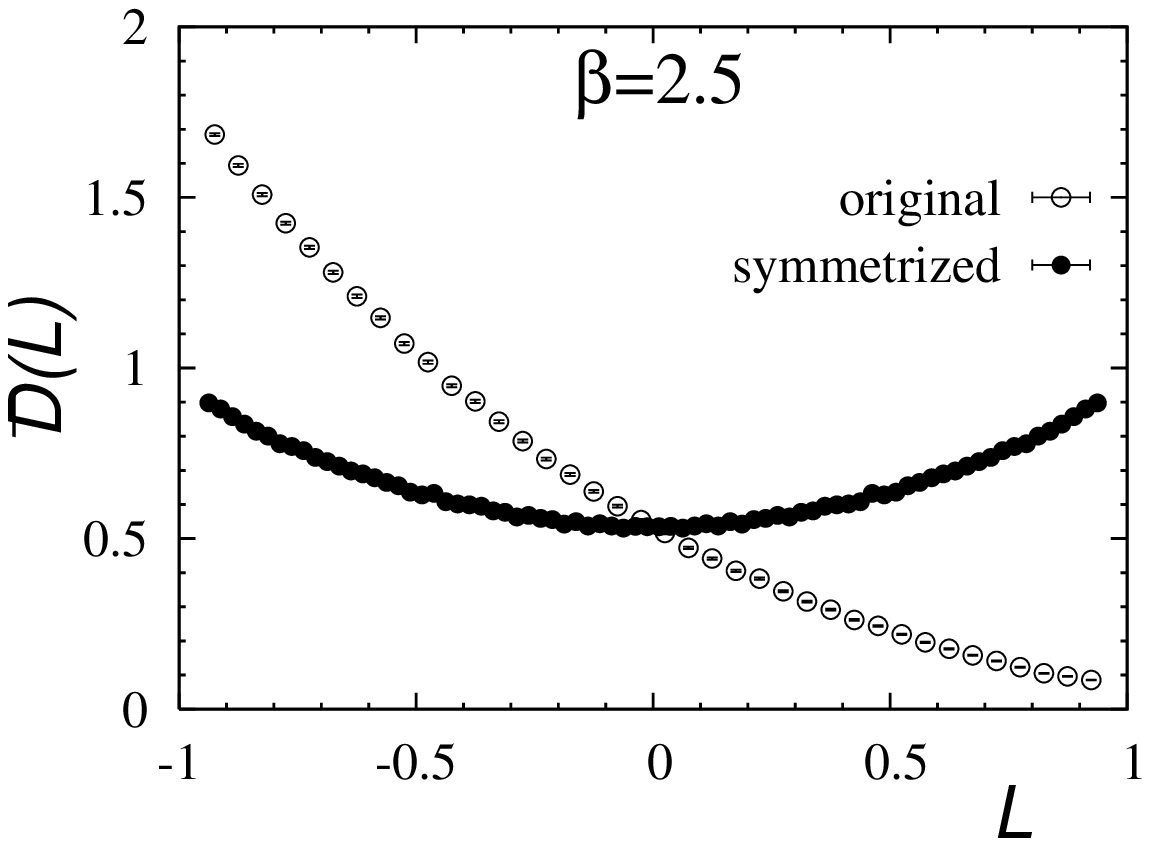} \\
(a) & (b)
\end{tabular}
\caption{The same as in Fig.~\ref{fig:typical:distributions} but for normalized
distributions ${\bar D}(L) = D(L)/\sqrt{1-L^2}$.}
\label{fig:typical:distributions:normalized}
\end{figure}

In the confinement phase, the fluctuations of the Polyakov line are
random and are therefore basically given by the symmetric Haar measure
$D_{\mathrm{Haar}}(L){\mathrm d} L = (2/\pi) \sqrt{1-L^2}\,{\mathrm d} L$.
In the deconfinement phase, the distribution is asymmetric
and, consequently, far from flat. The distributions
normalized by the Haar measure are shown in
Fig.~\ref{fig:typical:distributions:normalized}.

\begin{figure}[htb]
\begin{tabular}{cc}
\includegraphics[width=65mm,angle=0,clip=true]{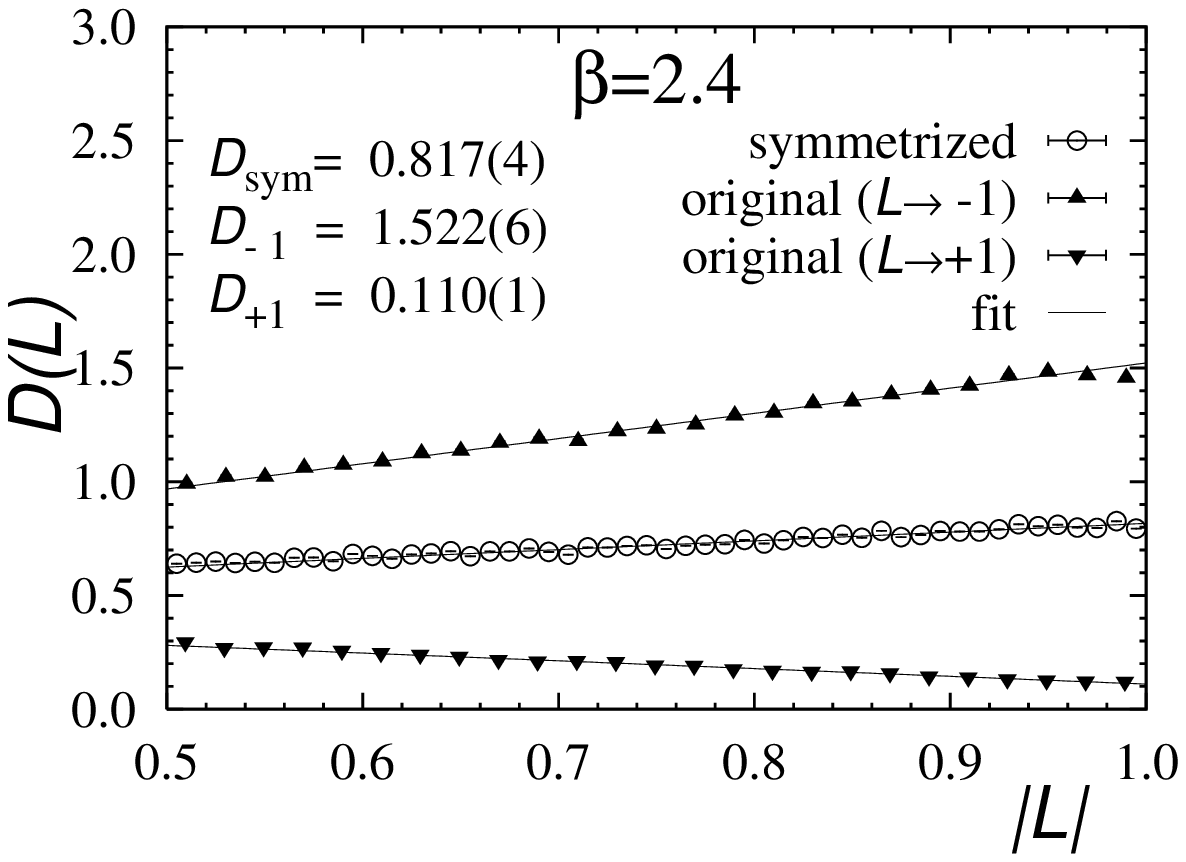} &
\includegraphics[width=65mm,angle=0,clip=true]{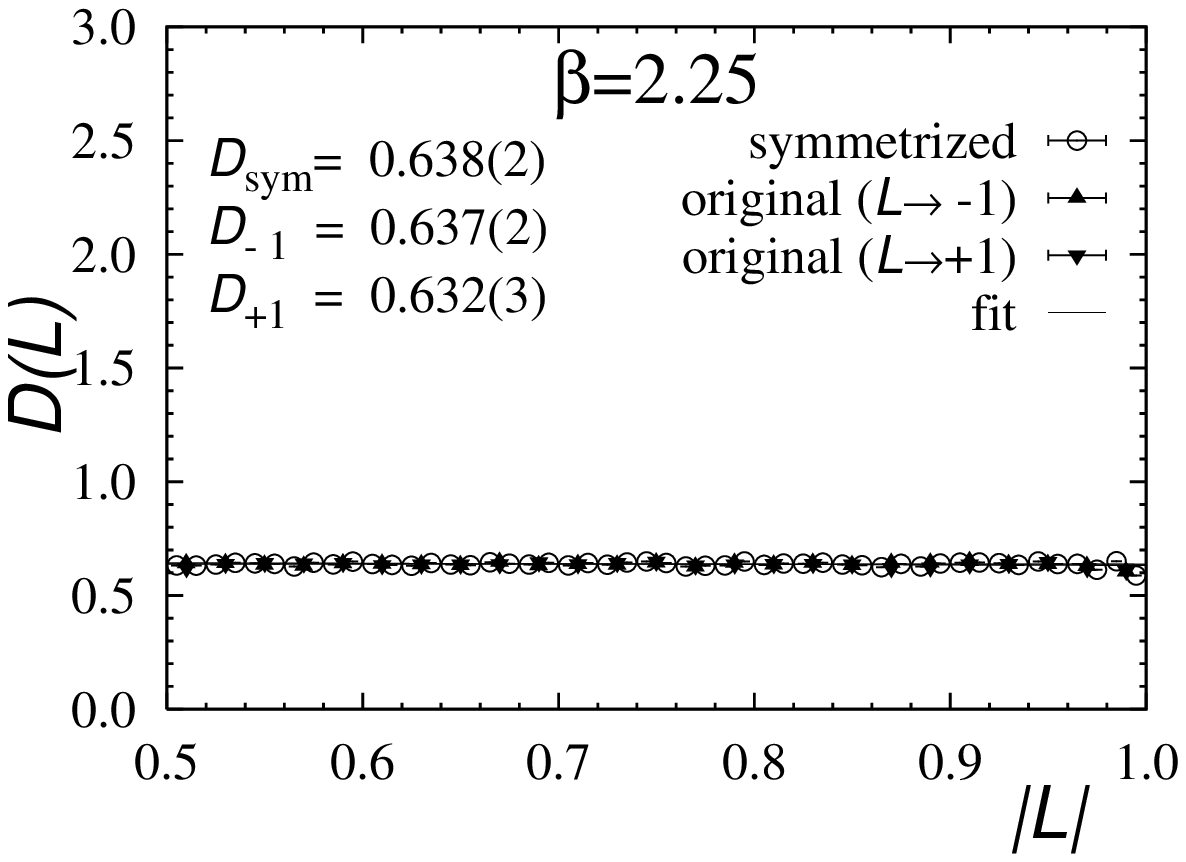} \\
(a) & (b)
\end{tabular}
\caption{Fits~\eq{eq:fitting} of the normalized distributions
${\bar D}(L) = D(L)/\sqrt{1-L^2}$ in the confinement (a) and
deconfinement (b) phases at both sides $L\to \pm 1$ of the distributions.}
\label{fig:symm:polyakov:fits}
\end{figure}

The qualitative behavior of the normalized distributions of the
Polyakov line are very interesting to us because the Haar factor
appears explicitly in the definition of
density of the static hedgehog lines~\eq{eq:object:density}. The density of
hedgehog lines thus
explicitly excludes the effect of the Haar measure. The examples
in Fig.~\ref{fig:typical:distributions:normalized} show that the
normalized distribution of the Polyakov line is finite at the
extremal end $L\to \pm 1$. To obtain an insight into the
behavior of the extremal ends, we fit the Polyakov distributions
by the simple linear function \beqn {\bar D}^{\mathrm{fit}}(L) =
D_z + C_z (1 - z\, L)\,, \quad \mbox{as} \quad L\to z = \pm 1\,,
\label{eq:fitting} \eeqn where $D_z$ and $C_z$ are the fitting
parameters labeled by the center element $z$. Examples of
fits~\eq{eq:fitting} are shown in
Fig.~\ref{fig:symm:polyakov:fits}.

We also show the fits of the symmetrized distributions. As can be seen from
these figures, the fits work very well. The extreme values of the
distributions $D_{-1} \neq D_{+1}$ are clearly asymmetric in the
deconfinement phase,
$D_{\pm 1}\neq D_{\mathrm{sym}}\equiv\frac{1}{2}(D_{-1}+D_{+1})$
and are symmetric, $D_{\pm1} = D_{\mathrm{sym}}$, in the confinement phase.

\begin{figure}[!htb]
\begin{tabular}{cc}
\includegraphics[width=65mm,angle=0,clip=true]{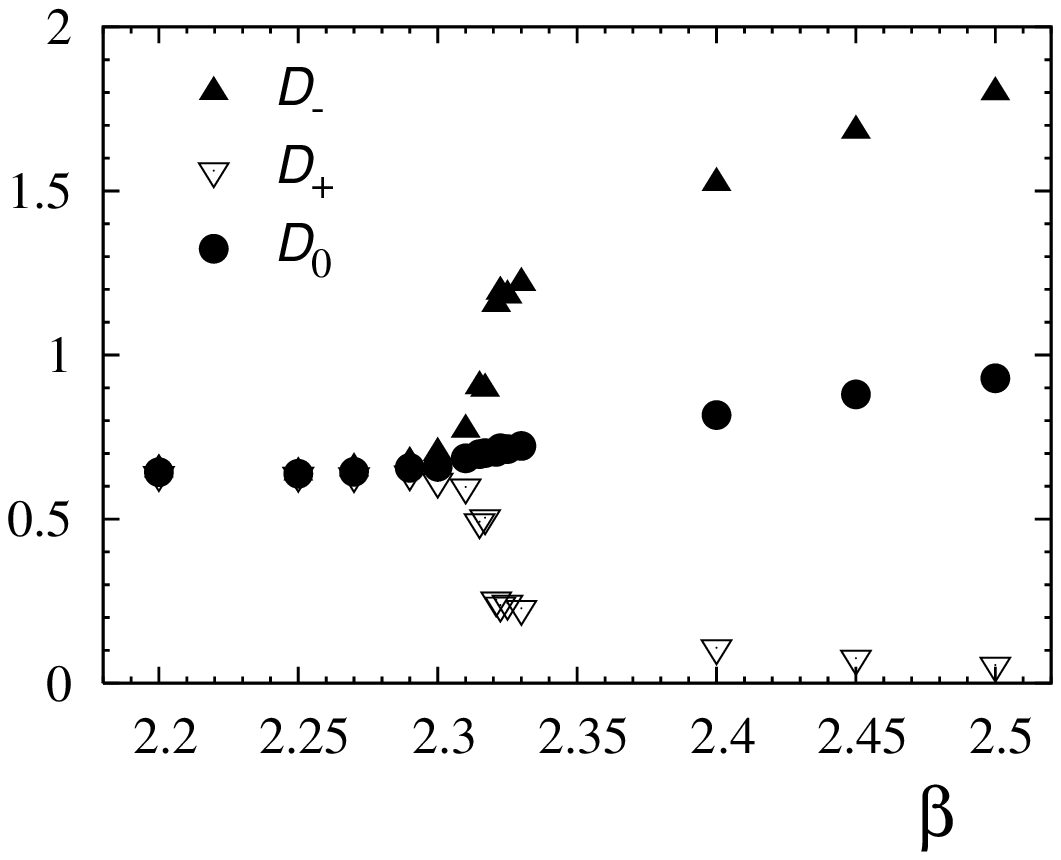} &
\includegraphics[width=65mm,angle=0,clip=true]{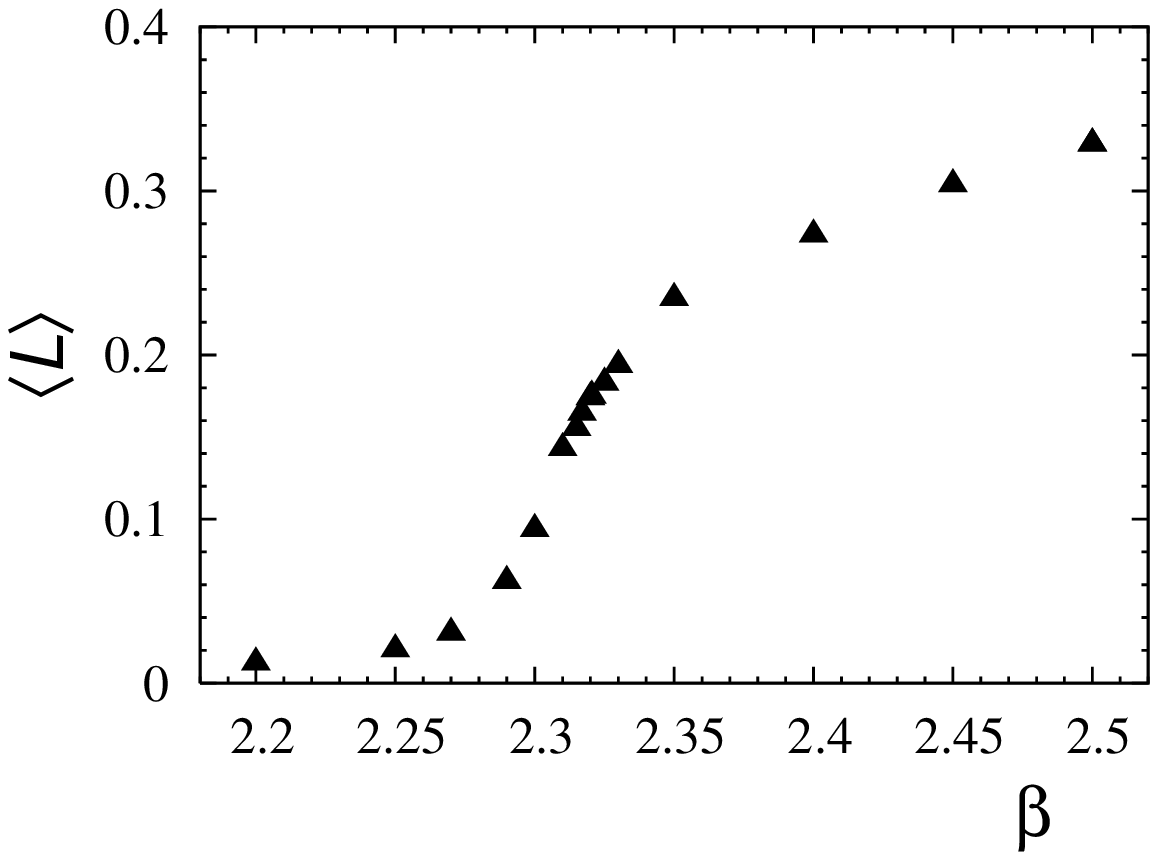} \\
(a) & (b)
\end{tabular}
\caption{(a) The extreme values $D_z$ of the normalized distributions
of the Polyakov line. (b) The expectation value of the Polyakov line
$|\langle L \rangle|$ as a function of the coupling $\beta$.
The critical temperature $T_c$ corresponds to $\beta_c \approx 2.30$,
and the deconfinement phase is realized at $\beta>\beta_c$
(equivalent to $T>T_c$).}
\label{fig:asymptotic}
\end{figure}

The extreme values $D_z$ of the normalized distributions $\bar D(L)$
are shown in Fig.~\ref{fig:asymptotic}(a)
as functions of the lattice coupling $\beta$.
In the confinement phase ($\beta < \beta_c$), the quantities
$D_z(\beta)$ are barely dependent on temperature ($\beta$).
But as the temperature increases further, the system passes
through the second-order phase transition, and the extreme
quantities $D_z(\beta)$ with $z=\pm 1$ show a rapid change at the
deconfinement point $\beta \approx \beta_c$. It can be clearly
seen that the values $D_{-1}$ become enhanced while the quantity
$D_{+1}$ becomes suppressed in the deconfinement phase. This
corresponds the system residing in the $z=-1$
vacuum\footnote{For illustrative purposes, we always show
the cases $D_{-1}(\beta) > D_{+1}(\beta)$ in the deconfinement
phase. In a finite volume, the system jumps from the $z=+1$
vacuum to the $z=-1$ vacuum and back. As the volume of the system
increases these jumps are suppressed.}. For comparison, the
expectation value of the Polyakov line as a function of $\beta$ is shown
in Fig.~\ref{fig:asymptotic}(b).

\begin{figure}[!htb]
\begin{tabular}{cc}
\includegraphics[width=65mm,angle=0,clip=true]{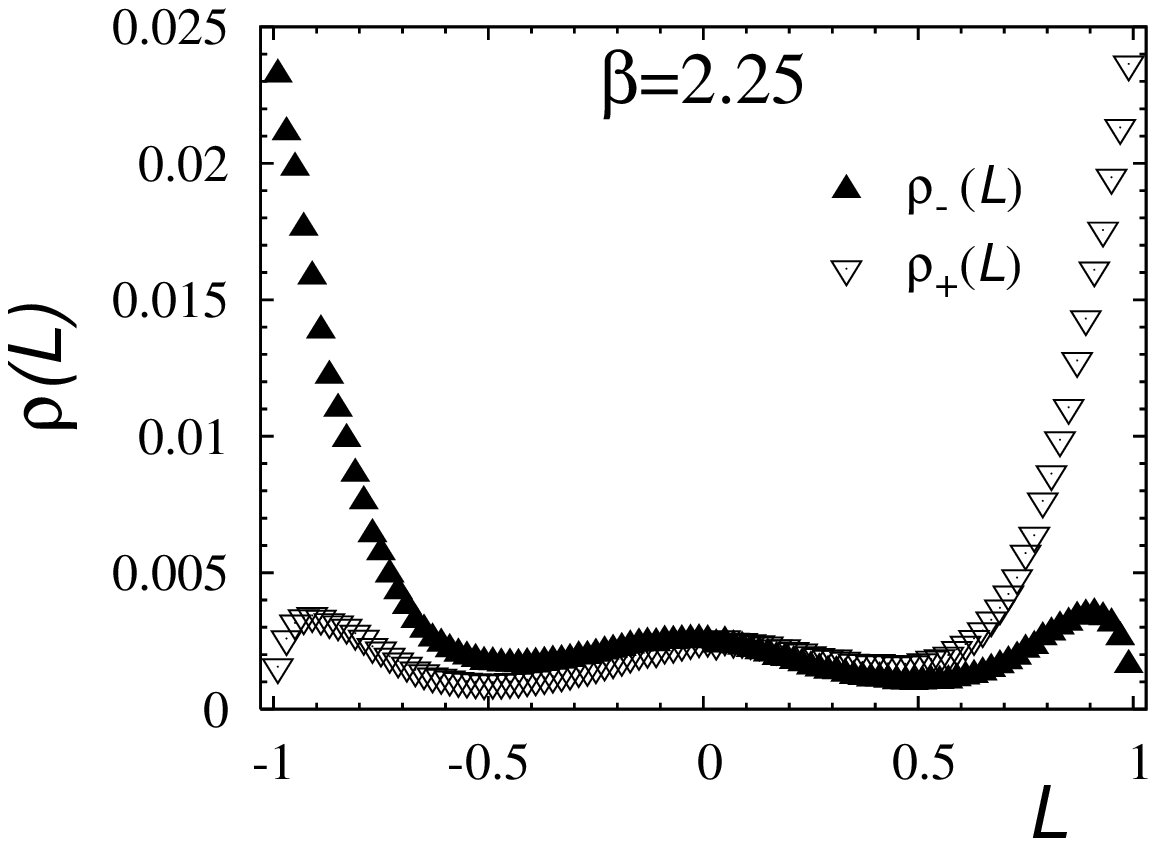} &
\includegraphics[width=65mm,angle=0,clip=true]{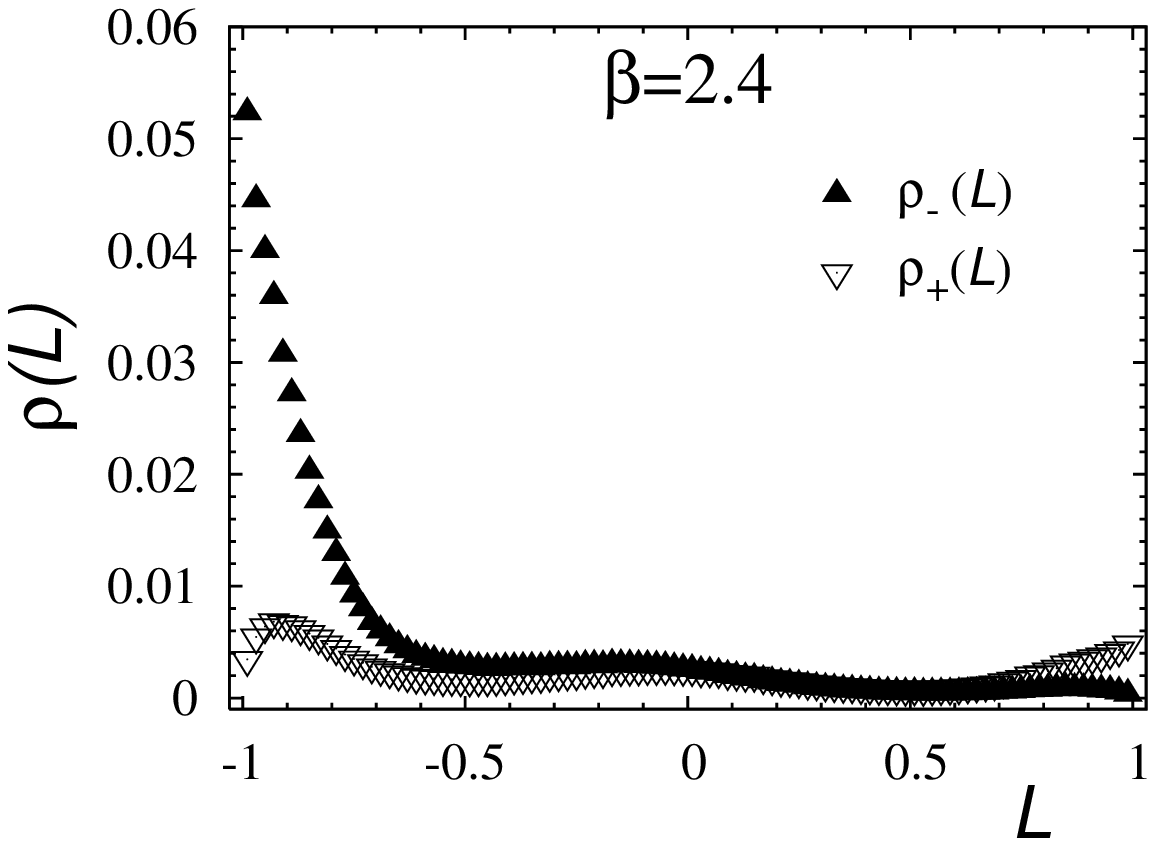} \\
(a) & (b)
\end{tabular}
\caption{Distributions~\eq{eq:rho:distribution} as functions of the
Polyakov line $L$ in the confinement (a) and deconfinement
(b) phases.}
\label{fig:rho:L}
\end{figure}

Distributions~\eq{eq:rho:distribution} corresponding to the
contributions~$\varrho_{\pm}(L)$ of the negative and positive
hedgehog lines~\eq{eq:object:density:evaluate:2} to total hedgehog line
density~\eq{eq:object:density:evaluate:1} are shown in
Fig.~\ref{fig:rho:L}
in the confinement and deconfinement phases.
They indicate that the averaged
quantities~$\varrho_{\pm}(L)$ are not symmetric functions of $L$
both in the confinement and in the deconfinement phases:
$\varrho_{\pm}(L)\neq \varrho_{\pm}(-L)$. This observation
stresses the importance of $d_{\pm}$
factors~\eq{eq:object:density:evaluate:2} in the definition of the
hedgehog line density. It can also be seen that the density distributions
in the confinement phase are cross-symmetric in the sense that
$\rho_{\pm}(L) = \varrho_{\mp}(- L)$. This cross-symmetry,
which is a direct consequence of the ${\mathbb Z}_2$ center symmetry,
is violated in the deconfinement phase.

\begin{figure}[!htb]
\centerline{\includegraphics[width=100mm,angle=0,clip=true]{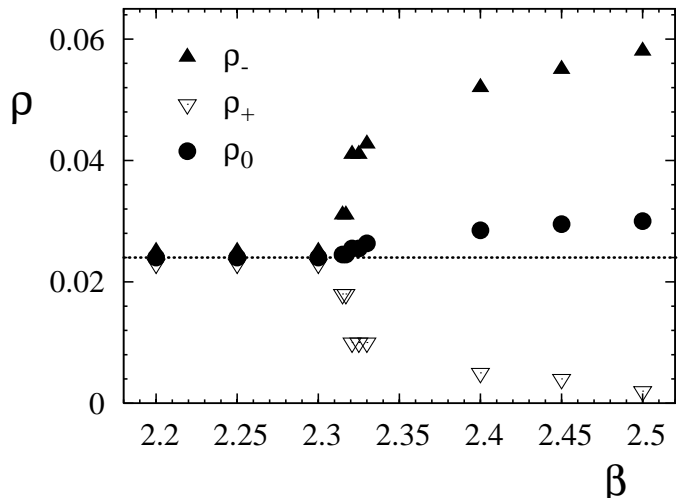}}
\caption{The densities $\rho_z$ of static hedgehogs corresponding to the
positive and negative central charges $z=+1$ and $z=-1$
as a function of $\beta$. The vacuum resides at the negative
center value $z=-1$. The dotted horizontal line marks the value
$\rho_{\mathrm{conf}}$ of the densities in the confinement phase.}
\label{fig:rho}
\end{figure}

Finally, the density of static hedgehog lines is given by the
extreme values $L \to \pm 1$, Eq.~\eq{eq:rho:average}, of
distributions~\eq{eq:rho:distribution}. To obtain these densities,
we used a linear-fit function similar to Eq.~\eq{eq:fitting}. The
densities $\rho_-$ and $\rho_+$ of the hedgehog lines
with the respective center charges $z=-1$ and $z=+1$ are shown in
Fig.~\ref{fig:rho}
as a function of the gauge coupling $\beta$.
A rapid increase in the density of $z{=}-1$ hedgehog lines, which correspond
to the value of the center group closest the broken vacuum state,
can be clearly observed. The hedgehog lines with the center charge $z=+1$
are far from the broken vacuum and are consequently suppressed
in the infinitely high temperature limit.
The symmetrized density, $\rho_0 = \frac{1}{2}(\rho_- + \rho_+)$,
shows a less pronounced, but still noticeable, increase
as the temperature increases in the deconfinement phase. In
the confinement phase, all densities take the same value
$\rho_{\mathrm{conf}} \approx {\mathrm{const}}$ and are
independent of the temperature ($\beta$) within error bars. Our results
thus indicate that the additively normalized density
\beqn
\delta
\rho_z(T) = \rho_z(T) - \rho(T=0)\,,\qquad z=\pm 1\,,
\eeqn
(as well as the corresponding symmetrized quantity $\delta \rho_0$)
is an order parameter of the deconfinement phase transition.

Note that in the lattice units the density $\rho_\pm$ of the hedgehog
lines - shown in Figure~\ref{fig:rho} -- is more than by an order of magnitude
smaller than the extreme values $D_z$ of the normalized distributions of the
Polyakov line, Figure~\ref{fig:asymptotic}(a). The difference comes
form the $d_\pm$ factors~\eq{eq:object:density:evaluate:3}, which enter
the definition of the density $\rho_\pm$ given in Eqs.~\eq{eq:rho:distribution}
and \eq{eq:rho:average}.

We have also verified that the dependence of our results on the
bin size of the distributions (shown in Fig.~\ref{fig:rho:L})
is very weak. For example, at $\beta=2.33$ corresponding to the
deconfinement phase, for one of the
densities, the values $\rho_+ =0.0427(17)$, $0.0455(29)$, and $0.0468(41)$,
correspond to the respective bin numbers $N_L = 100$, 200, and 300.
A similar weak dependence on the number of bins is
observed for $\rho_-$. The densities at other values
of $\beta$ show the same behavior.

It is important to note the following. One could expect from the
very beginning that the static hedgehog lines -- being defined by the
center elements of the gauge group -- should be dense in the
high-temperature phase since at very high temperatures the
Polyakov line approaches a center element. However, this argument
is not valid because, for example, it also leads to the erroneous
conclusion that the hedgehog lines should be dense in the trivial
vacuum, $A^a_\mu = 0$. In fact, arguments of
Ref.~\cite{ref:objects} indicate that the trivial vacuum does not
possess the hedgehog lines at all. Technically, vanishing of the
density of the hedgehog lines in the trivial vacuum is guaranteed by the
determinant in Eq.~\eq{eq:object:density}. Thus, it is not clear
{\it ad hoc} should the density of the hedgehog lines be high in the
deconfinement phase or not. Consequently, the enhancement of the
density of the hedgehog lines in the high temperature phase,
Figure~\ref{fig:rho}, is a non-trivial dynamical fact.

\section{Conclusions}

We have given analytic and numerical arguments supporting the
suggestion~\cite{ref:objects} that the gauge-invariant
hedgehoglike structures in Wilson loops ("the hedgehog loops") are
physically interesting degrees of freedom in Yang--Mills theory. The
hedgehog loops can be defined self-consistently in terms of the
Wilson loop variables. The hedgehog loops are
associated with the (untraced) Wilson loops that take values in the
center of the gauge group, $z \in {\mathbb{Z}}_N$. In the SU(2)
Yang--Mills theory, the hedgehog loops are characterized by two quantized
numbers: the center charge $z = \pm 1 \in {\mathbb{Z}}_2$ and the
winding number $n \in {\mathbb{Z}}$. Because the hedgehog
loops are closed by construction, it seems natural to
suggest that the corresponding charges $z$ and $n$ are conserved.
In fact, the conservation of the center charge $z$ was proved
in~\cite{ref:objects} and was discussed here.
The conservation of the winding number $n$ along the hedgehog loop
and a stability analysis of the hedgehog configurations in particular models
may require an additional analysis.

{}From a general standpoint, it is clear that the hedgehog loops
should be related to the confinement properties of the system
because they are formulated in terms of the Wilson loops, which in
turn are intimately related to color confinement in the Yang--Mills
theory. The gauge invariance of hedgehog loops provides an additional argument
in favor of their importance for gauge-invariant phenomena, like the color
confinement. Unfortunately, the nonlocal formulation of the
hedgehogs makes it difficult to locate these objects directly in
separate configurations of the gluon fields. On the other hand,
this difficulty can be circumvented by noticing that the statistical
properties of hedgehogs can be studied using particular distributions
corresponding to these objects with trajectories of predefined shapes.
These distributions should provide information about densities,
correlation properties, {\it etc.}, of these objects.

In this paper, we have demonstrated the success of the proposed numerical
approach by studying the properties of static hedgehog lines in the
finite-temperature SU(2) gauge theory. By definition, static (or, "thermal")
hedgehog line correspond to extremal ({\it i.e.}, center) values of the Polyakov
line variables, which
should naturally arise in an effective model of the Polyakov lines~\cite{ref:Pisarski1}.
This model is used to describe features of the finite-temperature QCD transition
expected to be realized in the high-energy heavy-ion
collisions~\cite{ref:Pisarski1,ref:Pisarski2}. We showed numerically
that the density of thermal hedgehogs changes rapidly
as the temperature increases in the deconfinement phase. We observed
a substantial increase (decrease) of the density of the thermal hedgehog
lines with the central charge $z$ closest to (furthest from) the
value of the Polyakov line in the centrally broken vacuum of the
deconfinement phase. We thus found evidence that the hedgehog line
density (additively normalized to zero at zero temperature) is an
order parameter of the deconfinement phase transition. We conclude
that hedgehogs in Wilson loops (in particular, in Polyakov lines) may be
relevant degrees of freedom of the Yang--Mills vacuum.

\section*{Acknowledgments}
The authors are grateful to M.\,I.~Polikarpov for participation in
an early stage of this project. M.\,N.\,Ch. is grateful to
F.~Bruckmann and G.~Burgio for the useful discussions and
suggestions. The authors are supported by a STINT Institutional
grant IG2004-2 025, by the grants RFBR 04-02-16079, RFBR
05-02-16306, DFG 436 RUS 113/739/0, MK-4019.2004.2., by the
``Dynasty'' foundation and ICFPM. M.\,N.\,Ch. is thankful to the
members of Department of Theoretical Physics of Uppsala University
for the kind hospitality and stimulating environment.

\end{document}